\documentclass{agnsymp}      
\usepackage{psfig,graphics}
\def\ros{{\sl ROSAT }}

\def\arcmin{\ifmmode ^{\prime}\else$^{\prime}$\fi}
\def\arcsec{\ifmmode ^{\prime\prime}\else$^{\prime\prime}$\fi}
\def\approxlt{\mathrel{\hbox{\rlap{\lower.55ex \hbox {$\sim$}}
        \kern-.3em \raise.4ex \hbox{$<$}}}}
\def\approxgt{\mathrel{\hbox{\rlap{\lower.55ex \hbox {$\sim$}}
        \kern-.3em \raise.4ex \hbox{$>$}}}}

\begin{document}
\title{Warm Absorbers in Active Galactic Nuclei} 
\author{Stefanie Komossa} 
\institute{Max--Planck--Institut f\"ur extraterrestrische Physik,
Postfach 1603, 85740 Garching, Germany \\
 {\sl email: skomossa@xray.mpe.mpg.de}  }
\authorrunning{St. Komossa}
\titlerunning{Warm Absorbers in AGN}
\maketitle

\begin{abstract}

We first provide a review of the properties of warm absorbers 
concentrating on what we have learned from {\sl ROSAT} and {\sl ASCA}.
This includes dusty and dust-free warm absorbers, non-X-ray
emission and absorption features of warm absorbers, 
and the possible warm absorber interpretation 
of the peculiar 1.1 keV features. 

We then discuss facets of warm absorbers by a more detailed
investigation of individual objects: 
In a first part, we discuss several candidates for dusty warm absorbers.
In a second part, we review and extend our earlier study of a possible
relation between warm absorber and CLR in NGC\,4051, and confirm that
both components are of different origin (the {\em observed} coronal
lines are {\em under}predicted by the models, the warm absorber 
is too highly ionized). We then suggest that a potential overprediction
of these lines in more lowly ionized absorbers can be avoided 
if these warm absorbers are dusty. 
In a third part, we present first results of an analysis of a
deep {\sl ROSAT} PSPC observation
of the quasar MR2251--178, the first one discovered to host
a warm absorber. 
Finally, we summarize our scrutiny under which conditions
a warm absorber could account for the dramatic
spectral variability of the Narrow-line Seyfert\,1 galaxy RXJ0134-4258.

\end{abstract}

\section{Introduction}

Warm absorbers (WAs) are an important new diagnostic of the 
physical conditions  
within the central regions of active galaxies.
They have been observed in $\sim$50\% of the well-studied Seyfert galaxies
and have also been detected in quite a number of 
Narrow-line Seyfert\,1 galaxies. 
So far, they revealed their existence mainly in
the soft X-ray spectral region where they show up
by their characteristic metal absorption edges. 
The study of the ionized material provides a wealth of information
about the nature of the warm absorber itself, its relation
to other components of the active nucleus, and the intrinsic
AGN X-ray spectral shape, and leads to a   
 better  understanding of AGN in general. 

The presence of an ionized absorber was first 
discovered based on {\sl Einstein} observations
of the quasar MR 2251-178 (Halpern 1984). 
With the improved spectral capabilities of {\sl ROSAT} and
{\sl ASCA}, many more warm absorbers have been 
found{\footnote{e.g., 
MCG\,6-30-15 (Nandra \& Pounds 1992, Fabian et al. 1994,
 Otani et al. 1996b, Reynolds et al. 1997, Orr et al. 1997, and
references therein),
NGC\,3783 (Turner et al. 1993, George et al. 1995, 1998c), 
3C\,212 (Mathur 1994),
3C\,351 (Fiore et al. 1993, Mathur et al. 1994, Nicastro et al. 1999b),
NGC\,3227 (Ptak et al. 1994, Komossa \& Fink 1997b, George et al. 1998b),
NGC\,4051 (e.g., Pounds et al. 1994, Mihara et al. 1994, McHardy et al. 1995,
Guainazzi et al. 1996, Komossa \& Fink 1997a, Guainazzi et al. 1998,
and references therein), 
NGC 4151 (Weaver et al. 1994, Warwick et al. 1995, 1996), 
IC 4329A (Cappi et al. 1996, Perola et al. 1999), 
NGC 3786 (Komossa \& Fink 1997c),   
NGC\,5548 (Nandra et al. 1993, Mathur et al. 1995, 1999, Done et al. 1995),  
NGC\,3516 (Kriss et al. 1996a, Mathur et al. 1996), 
Mrk\,290 (Turner et al. 1996), 
Mrk\,1298 (Wang et al. 1996, 1999, Komossa \& Fink 1997d, Komossa \& Meerschweinchen 2000), 
IRAS\,13349+2438 (Brandt et al. 1996, 1997, Komossa et al. 1999b, Siebert et al. 1999),  
IRAS\,17020+4544 (Leighly et al. 1997, Komossa \& Bade 1998), 
PG\,1114+445 (George et al. 1997),
4C +74.26 (Brinkmann et al. 1998, Komossa \& Fink 1998);
see Reynolds (1997) and George et al. (1998a)
for a systematic study of WAs with {\sl ASCA} }},
and theoretical modeling of
plasma under `warm absorber' conditions   
has been pushed forward{\footnote{e.g., Halpern 1984, Krolik \& Kallman 1984, 
Netzer 1993, 1996, Krolik \& Kriss 1995, Murray et al. 1995, 
Reynolds \& Fabian 1995, Komossa \& Bade 1998, Nicastro
et al. 1999a}}.  
 
Below, we first provide a review on
what is known on warm absorbers 
after the `{\sl ROSAT} and {\sl ASCA} epoch' 
of observations and theoretical modeling (Sect. 2).
We then discuss in more detail the possibility that
several (but not all) warm absorbers contain dust (Sect. 3).
In Sect. 4 we investigate the possible relation of
WA and optical/UV emission line regions with focus on NGC\,4051. 
Sect. 5 is concerned with the warm absorber in MR\,2251-178. 
In Sect. 7 
we explore whether and under which conditions a 
warm absorber provides an explanation for the
dramatic X-ray
spectral variability of RX\,J0134-4258.

\section{Basic properties of warm absorbers: dusty WAs, dust-free WAs, 
         peculiar 1.1\,keV features}

\subsection{Warm absorber models and definitions}

The active nucleus is thought to contain 
an accreting supermassive black hole (SMBH),
surrounded by the putative molecular torus and two emission 
line regions, the broad line region (BLR) and the narrow line region (NLR).
The physical properties of the two emission line regions
are quite well known, since a wealth of information can be extracted
from emission line intensities, line profiles, line shifts, 
and line variability (see Peterson 1993 for a review). 

Concerning the nature and location of the warm absorber, 
several different models have been
suggested: (i) a relation of the WA to the BLR  
(a high-density component of the inner BLR,  
a BLR confining medium, winds from bloated stars, or a matter bounded BLR component;
see, e.g., Reynolds \& Fabian 1995, Shields et al. 1995, Netzer 1996),
(ii) an accretion disk wind (e.g., K\"onigl \& Kartje 1994, Murray et al. 1995),
(iii) a relation to the torus (e.g., Reynolds 1997, Komossa \& Fink 1997b), 
or (iv) a relation
to the NLR (see, e.g., the two-component WA model of 
Otani et al. 1996b). The reason for the
large variety of models discussed is that not all 
physical properties of the warm absorber (its density $n$, 
column density $N_{\rm w}$, covering factor $\eta$ = $\omega/4\pi$,
distance $r$ from the nucleus, elemental abundances $Z$,
its velocity field, and the shape of the illuminating continuum) 
can be directly
determined from X-ray spectral fits, but only
certain combinations of these parameters, defined below.       
 
The two most important quantities are the column density
$N_{\rm w}$ and ionization parameter $U$ of the warm absorber.  
$N_{\rm w}$ is given by  
    \begin {equation}
    N_{\rm w} = \int {n_{\rm H}} {\rm{d}}x ~~,
    \end {equation}
where $n_{\rm H}$ is the total hydrogen density, and $x$ 
the thickness of the absorber.      
The ionization parameter $U$ is defined as 
    \begin {equation}
     U = {Q \over {4\pi{r}^{2}n_{\rm H}c}}~~.
    \end {equation}
Here, $Q$ relates to the spectral energy distribution (SED)
which illuminates the warm absorber, and is the number rate of photons 
above the Lyman limit{\footnote{In relation
to warm absorbers, some authors use an `X-ray ionization parameter',
counting only X-ray photons, because the ionization state of the
warm material is dominated by the soft X-ray, not the EUV-part
of the spectrum.}} ($\nu_0 = 10^{15.512}$ Hz), given by  
    \begin {equation}
    Q = 4\pi{d}^{2} \int\limits_{\nu_0}^{\infty} { f_{\nu} \over {h\nu}} \rm{d}\nu~~,
    \end {equation}
where $d$ is the distance between observer and warm absorber. 
The mass of the warm absorber is approximately given by 
    \begin{equation}
    M_{\rm wa} \simeq 4 \pi r^2 N_{\rm w} m_{\rm p} \eta  =  0.02 r_{16}^2 N_{22} \eta~ \rm{M_\odot}~~,
    \end{equation}
with $r = 10^{16}r_{16}$ cm and $N_{\rm w} = 10^{22}N_{22}$ cm$^{-2}$.  

From direct spectral fits, one first determines $U$ and $N_{\rm w}$.
If the intrinsic continuum turns out to be variable, or if there is evidence 
that the warm material contains dust, important constraints on the density
and distance of the absorber from the nucleus can be obtained
from reaction-timescale arguments, and dust survival arguments, respectively.   
In particular, the recombination timescale $t_{\rm rec}$ of a warm
absorber is given by
    \begin{equation}
    n_{\rm{e}} \approx {1\over t_{\rm{rec}}}~{n_{\rm{i}}\over n_{\rm{i+1}}}~{1\over A}({T\over 10^4})^{X} 
    \end{equation}
(Krolik \& Kriss 1995) where 
$n_{\rm{i}}/n_{\rm{i+1}}$ is the ion abundance ratio of the metal ions
dominating the cooling of the gas,
$n_{\rm{e}}$ the electron density,
and the last term is the
corresponding recombination rate coefficient $\alpha_{\rm{i+1,i}}^{-1}$ 
(Shull \& Van Steenberg 1982; for coefficients
$A$, $X$ see Aldrovandi \& Pequignot 1973). 
Available observations locate most warm absorbers outside the bulk
of the BLR.

\subsection{X-ray appearance of warm absorbers} 

  \begin{figure} [b]  
      \vbox{\psfig{figure=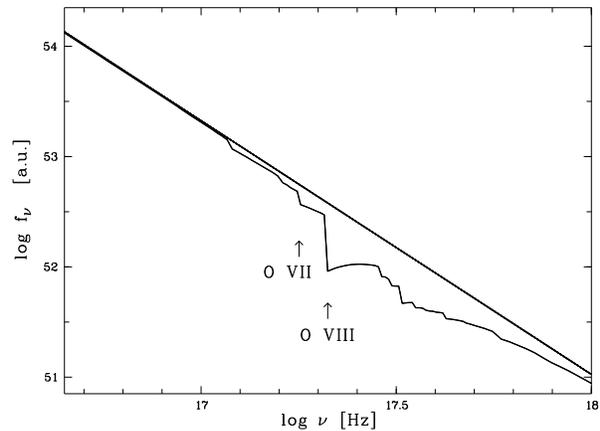,width=8cm,%
          bbllx=2.9cm,bblly=1.1cm,bburx=18.3cm,bbury=12.2cm,clip=}}\par
 \caption[]{X-ray absorption structure of a typical warm
absorber. The thin solid line corresponds to the intrinsic X-ray spectrum,
the thick solid line shows the spectrum after passage of the ionized
absorber. The oxygen OVII and OVIII edges are marked.  
}
\end{figure}

\subsubsection {Dust-free warm absorbers} 

So far, ionized absorbers have been
mainly observed at X-ray energies; therefore, most of
our knowledge on this material comes from the analysis
of X-ray spectra. 
Warm absorbers show up in the soft X-ray band 
by their characteristic X-ray absorption edges
which are created by K-shell photoionization of
highly ionized metal ions.
The warm material is thought 
to be photoionized by emission from the central continuum source in the
active nucleus (for a detailed model invoking shocks (Meier \& Demmig 1997)
in addition to photoionization
see Contini \& Viegas 1999; see also the discussion in Nicastro et al. 1999a). 
Its degree of ionization is higher than that of the bulk of the BLR
and the gas temperature is typically of order 10$^5$ K.
Fig. 1 shows the X-ray spectrum of a typical
warm absorber.
The most prominent
absorption edges are marked. 
In case of high covering factor of the warm absorber, 
X-ray emission and reflection becomes significant and will
partially fill up the edges (e.g., Netzer 1993 (his Fig. 1-3), 1996). 

\subsubsection {Dusty warm absorbers} 

Warm absorbers have been mostly modeled using the
codes {\em Cloudy} (Ferland 1993, Ferland et al. 1998) 
and {\em ION} (Netzer 1993, 1996).  
For several years, the ionized absorbers were assumed to be
dust-free; partly because they were observed in bright,
optically unreddened  Seyfert\,1
galaxies, and there was no necessity at all to suspect 
this highly ionized material to contain dust.  

The situation changed with the cases of 3C 212 (Mathur 1994), 
IRAS\,13349+2438 (Brandt et al. 1996), and NGC\,3227 (Komossa \& Fink 1996, 1997a): 
3C 212 is a quasar with a heavily reddened optical spectrum.
Its X-ray spectrum was successfully fit by several models
including a powerlaw plus excess
cold absorption (Elvis et al. 1994). Since other multi-wavelength
observations did not indicate a cold absorber of high column density
along the line of sight, Mathur (1994) suggested that the dust
that reddens the optical spectrum is mixed with a {\em warm}
absorber, instead. She also showed that this material can consistently 
account for the observed MgII UV absorption of 3C 212.   
In the cases of IRAS 13349 and NGC\,3277
the large column density $N_{\rm opt}$ of dusty
gas inferred from optical observations is inconsistent with 
the one ($N_{\rm x}$) derived from X-ray spectral fits 
in the sense that $N_{\rm x} \ll N_{\rm opt}$
(hereafter referred
to as `$N_{\rm opt}-N_{\rm x}$ discrepancy').   
The inconsistency disappears, if the dust is accompanied 
by {\em ionized} gas.  
Further candidates for dusty warm absorbers 
which show the same $N_{\rm opt}-N_{\rm x}$ discrepancy 
emerged quickly: 
NGC\,3786, MCG\,6-30-15, IRAS\,17020+4544, and possibly 4C74.26
(see Tab. 1). 

Warm absorber {\em models} taking into account the 
presence of dust were first presented
by Komossa \& Fink (1996, 1997a) and applied to the {\sl ROSAT} X-ray spectrum
of NGC\,4051 (which turned out to be likely dust-free, see Sect. 
4 below), followed by NGC\,3227, NGC\,3786 (which are well fit
by dusty warm absorbers; Komossa \& Fink 1997b,c), 
and MCG\,6-30-15 (which likely possess
a two-component WA one of which is dusty; Reynolds et al. 1997).    

In an alternative approach, Reynolds (1997) correlated the optical
depths of OVII and OVIII of a sample of Seyferts with the `steepness' of the UV continuum
which he 
took as reddening indicator. Except for two outlyers, most of
the objects of his sample show low reddening and weak OVII absorption,
whereas four objects show the opposite (high reddening, a larger
optical depth of OVII) and these latter are suggested to likely
host {\em dusty} warm absorbers. No such trend is found for OVIII.

  \begin{figure}[b]
      \vbox{\psfig{figure=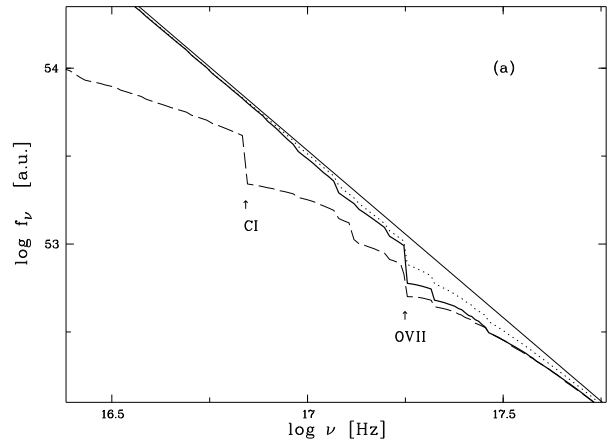,width=8.2cm,%
          bbllx=2.9cm,bblly=1.1cm,bburx=18.3cm,bbury=12.2cm,clip=}}\par
 \caption[]{Influence of dust mixed with the warm absorber on
       the X-ray absorption structure ($\log U = -0.5, \log N_{\rm w}$ = 21.6).
       The straight solid line corresponds to the unabsorbed
       intrinsic spectrum, the heavy solid line to a dust-free warm absorber (WA) with solar abundances,
       the dotted line to a dust-free WA with the same depleted abundances as in the dusty model,
       and the long-dashed line
       to the change in absorption structure after adding 
       dust to the WA (other model parameters are fixed).
       Clear changes in the absorption structure are revealed 
       for the models that include dust.
       The abscissa brackets the \ros energy range (0.1-2.4 keV); some prominent
       edges are marked. }
\end{figure}

\subsubsection{Peculiar 1.1 keV features}

Warm absorbers have also been observed in several
Narrow-line Seyfert\,1 galaxies (NLSy1s hereafter; e.g.,
Vaughan et al. 1999, Komossa \& Meerschweinchen 2000, and
references therein).
In addition to the `standard' warm absorbers with strong edges of OVII
and OVIII around 0.7-0.8 keV, peculiar absorption features
have been reported around $\sim$1 keV
(e.g., Brandt et al. 1994, Otani et al. 1996a, 
Ulrich-Demoulin \& Molendi 1996, Hayashida 1997,
Leighly et al. 1997, 
Matt 1998, Ulrich et al. 1999, Vaughan et al. 1999).  

Most interpretations of these feature focussed on warm
absorbers, albeit in very different ways. The suggested
explanations are:
(i) {\em Non-solar metal abundances}: Absorption around 1.1 keV 
could be due to Ne-K absorption but would require a rather peculiar
abundance ratio of Ne/O for the Ne absorption to be much stronger
than the O absorption. In addition, the strongest Ne edge
in most models is located at 1.36 keV at somewhat higher
energies than observed; Komossa \& Fink 1998, Ulrich et al. 1999.
Alternatively, it could be caused by Fe-L absorption (applied to PG1404+226
this requires an Fe overabundances of $>25 \times$ solar;
Ulrich et al. 1999).  
(ii) A warm absorber in {\em relativistic outflow}; the $\sim$1 keV
feature is then caused by highly blueshifted OVII or OVIII edges
(Leighly et al. 1997; see also Brandt et al. 1994 and Otani et al. 1996a).    
(iii) A sequence of {\em resonance absorption lines} in combination
with a steep X-ray spectrum (Nicastro et al. 1999c).  

It is important to keep in mind, though, that the observed features
seem to occur always close to the position where the
declining soft excess emission intersects the powerlaw
component, and the possibility remains that the absorption
feature is only mimicked by a more complex spectral
shape. Further, in the case of Ark\,564 the feature is better
described by an emission line than an absorption complex (Vaughan et al. 1999,
Turner et al. 1999).

\subsection{Relation to UV absorbers }

The `X-rays-only' approach to study warm absorbers
suffers from the   
still limited spectral resolution of current X-ray
detectors making measurements of, e.g., red/blueshifts
of the absorption edges difficult.
A powerful approach to constrain further the properties of warm absorbers
is to combine the X-ray with UV observations, and to search
for the signature of the warm absorbers simultaneously in
both spectral regions. 

The relation between UV and X-ray (cold) absorbers was examined 
by Ulrich (1988).  
Mathur and collaborators (e.g., Mathur 1994, Mathur
et al. 1994, Mathur et al. 1995) then
presented several good cases (3C 212, 3C351, NGC\,5548) 
where UV and X-ray warm absorber are
very likely one and the same component (see also deKool \& Meurs 1994
on PG\,1416-129). 
These absorbers are outflowing with velocities of
$\sim$1000-3000\,km/s and are located outside the BLR. 
Three further good candidates for a common UV--X-ray warm absorber
are the high-redshift quasar PKS 2351-154 
(Schartel et al. 1997), the Sy\,1 galaxy NGC\,3783 (Shields \& Hamann 1997),
and the quasar PG 1114+445 (Mathur et al. 1998).
Hamann et al. (1995a) reported the detection of NeVIII$\lambda$774
absorption in the quasar UM\,675, and noted that this gas could
act as X-ray warm absorber. 
Telfer et al. (1998) analyzed the
very highly ionized BAL system of the quasar SBS 1542+541 and
concluded that it is likely associated with an X-ray warm absorber.    
  
Sometimes, though, the UV absorption is rather complex and shows the presence
of different components with different degrees of ionization and/or velocities
(e.g., NGC 3516, Kriss et al. 1996b, Crenshaw et al. 1998; 
NGC\,5548, Mathur et al. 1999; NGC\,7469, Kriss et al. 1999),  
and only one component (e.g., NGC\,5548) or none (e.g, NGC\,7469)
may be identified 
with the X-ray warm absorber. 

In any case, the near one-to-one match of the presence of 
UV- and X-ray warm absorbers in a sample
of Seyfert\,1 galaxies (Crenshaw et al. 1999) suggests a 
common or related origin of UV and X-ray absorbing material. 

  \begin{figure} [t]  
      \vbox{\psfig{figure=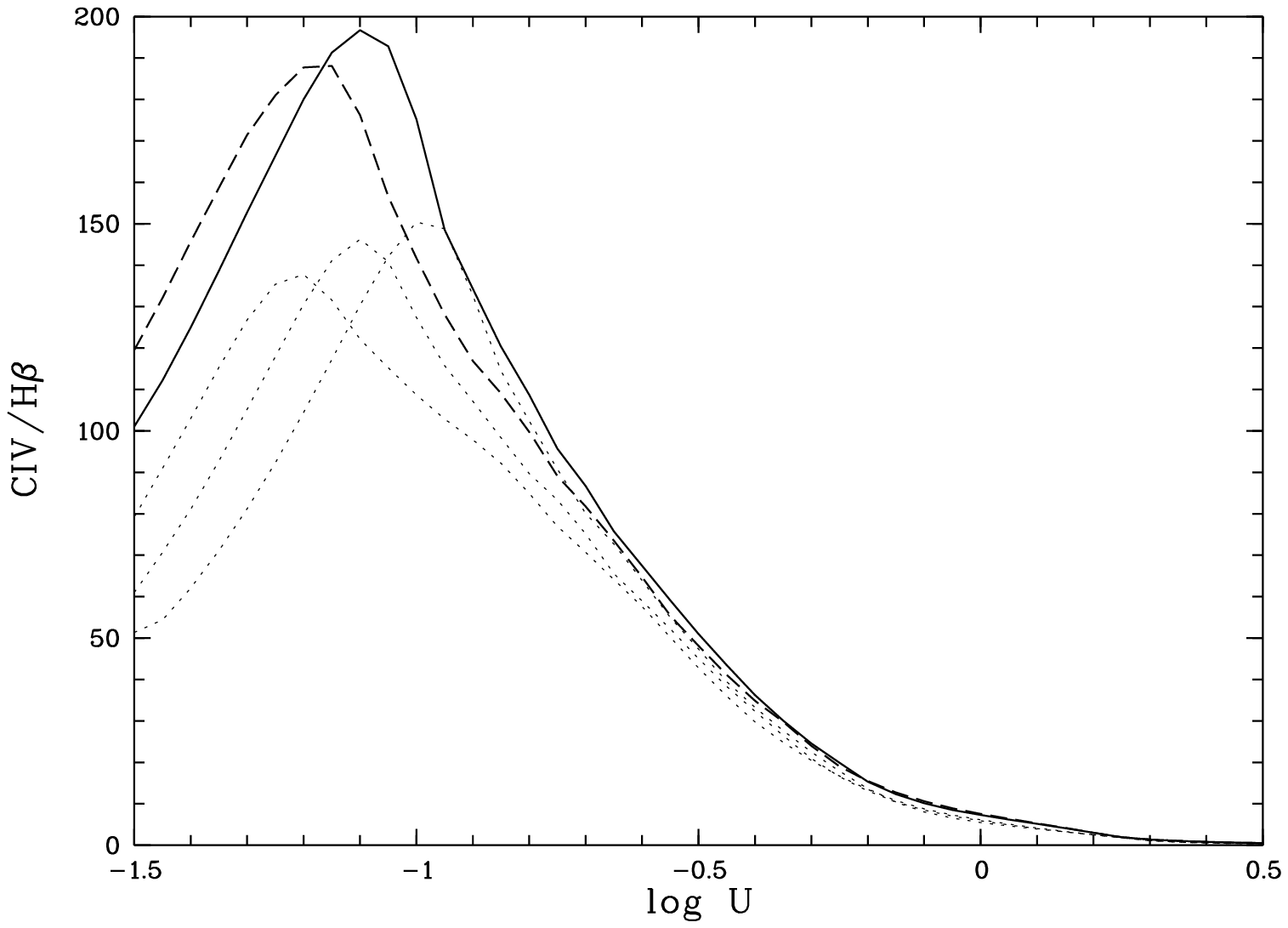,width=8cm,%
          bbllx=2.9cm,bblly=1.1cm,bburx=18.3cm,bbury=12.2cm,clip=}}\par
 \caption[C4]{Intrinsic warm absorber emission in CIV$\lambda$1549/H$\beta_{\rm WA}$,
in dependence of ionization parameter $U$ for different densities and
X-ray continuum shapes ($\alpha_{\rm EUV}$ was fixed to --1.4).
Solid line: spectrum with $\Gamma_{\rm x}$ = --1.9, log $N_{\rm w}$ = 21.5 und $\log n_{\rm H} = 9.5$;
dashed: $\Gamma_{\rm x}$ = --1.6;
dotted: $\Gamma_{\rm x}$ = --1.9, $\log n_{\rm H} = 7.0$ and, increasing from left to right) 
log $N_{\rm w}$ = 21.35, 21.5, 21.65.
}
\vspace*{0.4cm}
 \label{C4}
      \vbox{\psfig{figure=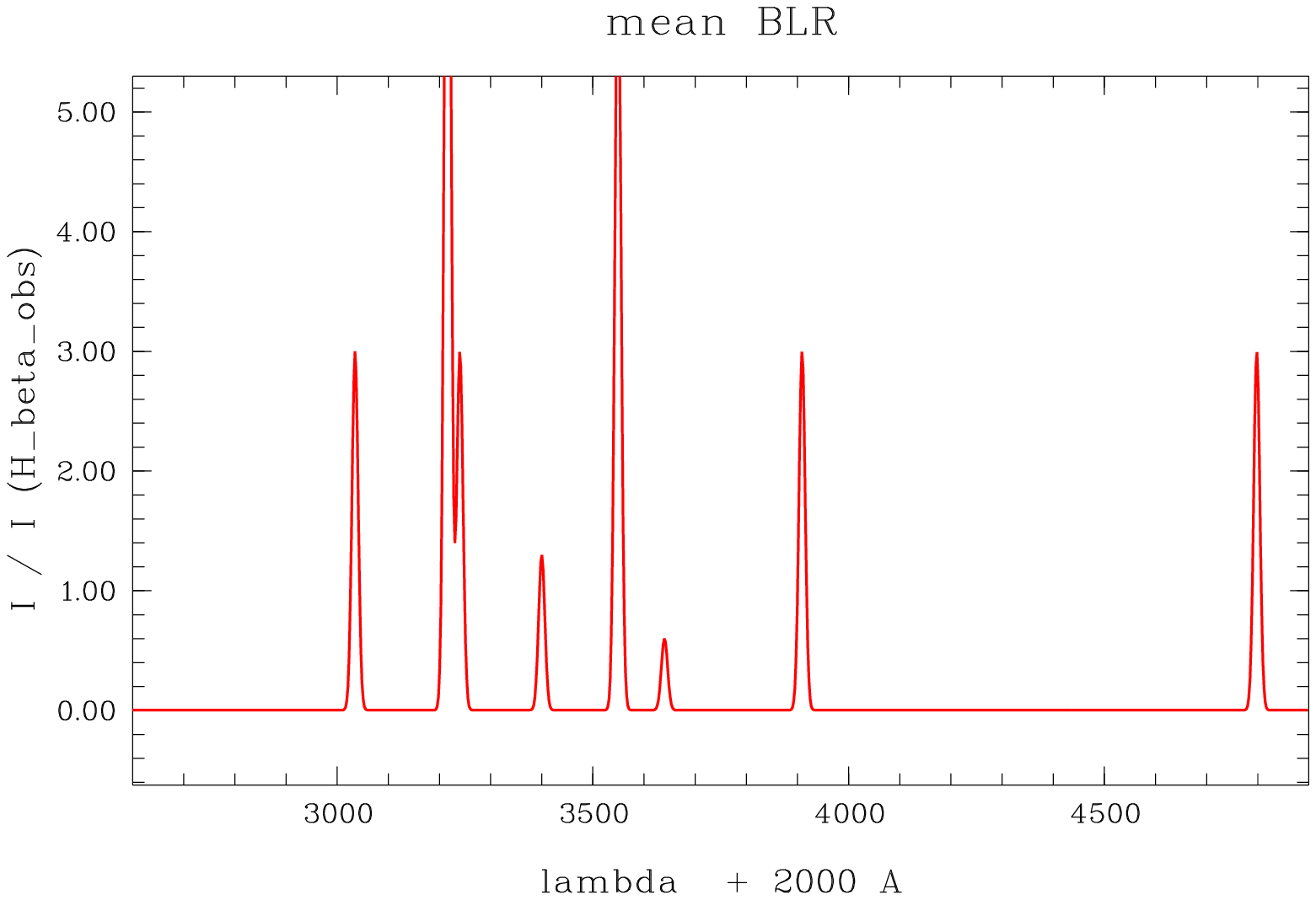,width=8.5cm,height=4.2cm,%
          bbllx=2.3cm,bblly=1.9cm,bburx=18.4cm,bbury=12.3cm,clip=}}\par
      \vbox{\psfig{figure=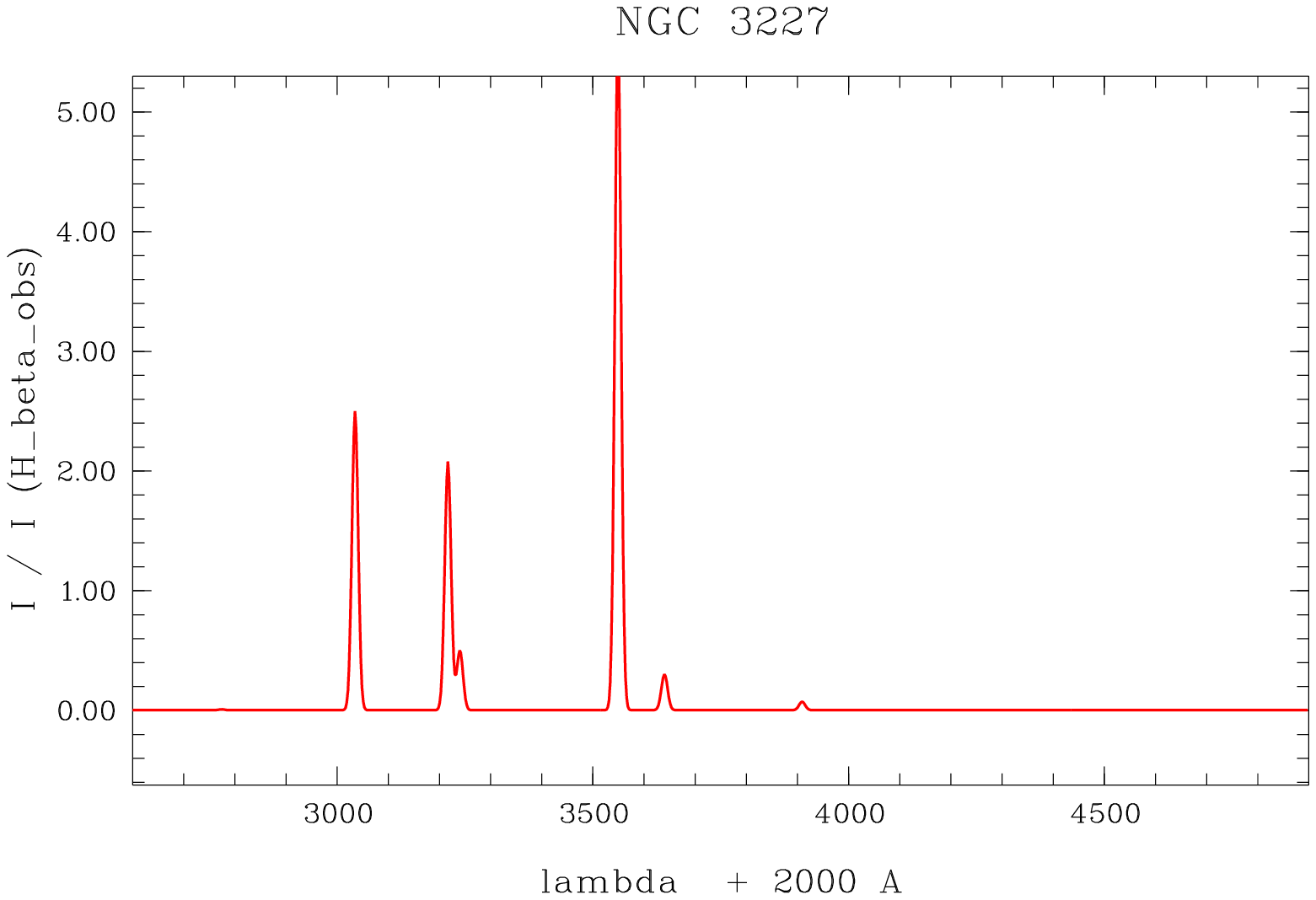,width=8.5cm,height=4.2cm,%
          bbllx=2.3cm,bblly=1.4cm,bburx=18.4cm,bbury=11.7cm,clip=}}\par
      \caption[emi]{Emission line spectra in the range 600 -- 2900 \AA~
predicted to arise from a warm absorber with $U = 0.1$ and
$\log N_{\rm w} = 21.5$ (bottom) compared
to the mean BLR spectrum (top; Netzer 1990). 
The y-axis gives 
the intensity ratio relative to H$\beta$.
In the bottom panel, the absorber-intrinsic emission lines 
were scaled to the mean observed H$\beta$ luminosity of NGC\,3227, 
to give an impression on the strengths of these lines and 
allow a judgement of their detectability.
}
\label{emi}
\end{figure}

\subsection{Non X-ray line emission from warm absorbers} 

Examples of UV-EUV lines that become strong
under warm absorber conditions were given in, e.g.,
Krolik \& Kriss 1995 (their Fig. 4), Shields et al. 1995,
Netzer 1996 (his Fig. 10), and
Komossa \& Fink 1997d\,(their Fig.\,6), 1997e. 

Observationally, the EUV emission line NeVIII$\lambda$774 was
detected in the spectra of high-redshift quasars (Hamann et al. 1995b,c, 1998)
and was suggested to originate from a warm absorber. (X-ray spectra
of these objects are not yet available for a check and more detailed
modeling).

A more detailed approach then was to select  
individual, well observed AGN and to investigate  
whether the physical conditions
in one of the observed optical-UV emission
line regions and the X-ray warm absorbers are identical,
such that the ionized X-ray absorber could be directly
identified with one of the known (high-ionization)
emission line regions. 
Mathur et al. (1994) presented detailed modeling 
of the BELR of the quasar 3C\,351 which also shows an
X-ray warm absorber, and concluded 
that high-ionization BELR and warm absorber of 3C\,351 are
not one and the same component.  
Komossa \& Fink (e.g., 1997a,b) studied the cases
of NGC\,4051 (summarized and extended in Sect. 4 below)
and NGC\,3227. They first derived
all warm absorber properties that can be
determined from X-ray spectral fits, then varied
the remaining free parameters (like metal abundances),  
predicted for each case 
the optical/UV line emission from the X-ray absorbing 
gas and compared with optical observations.
Again, they concluded that warm absorber and CLR in NGC\,4051,
and WA and emission-line components in NGC\,3227 
show different physical conditions.    
Reynolds et al. (1997) followed a similar approach
in application to MGC\,6-30-15
and find that [FeX] and [FeXIV]
could be explained by an (outer) warm absorber of 
covering factor $\sim$0.08, but that this would heavily underpredict
the observed [FeXI] emission. 
Contini \& Viegas (1999) presented a detailed 
multi-cloud model in which both, photoionization
and shocks contribute to the heating and ionization of the clouds,
and applied their model to NGC\,4051.
They find that the observed X-ray spectrum and the {\em NLR} line emission
can be successfully reproduced by an ensemble of clouds of different 
densities. [Predicted emission line intensities for coronal lines have not yet been
reported.]

\begin{table*}[t]
\caption { Summary of the candidates for {\em dusty} warm absorbers, listed in the chronological
           order they were suggested. The column ``sug.'' gives those references that
           proposed the presence of a {\em dusty} warm absorber, whereas the column ``mod.''
           lists the works that first {\em modelled} the X-ray data in terms of a dusty warm absorber. 
           Values of the ionization parameter $U$ (Eq. 2) given here and elsewhere in the text
           refer to a continuum spectrum with $\alpha_{\rm EUV}=-1.4$ (between Lyman-limit and 0.1 keV)
           and $\Gamma_{\rm x}$ as listed. }
  \begin{tabular}{lllcllll}
\noalign{\smallskip}
  \hline
      \noalign{\smallskip}
   object  & type & \multicolumn{3}{l}{~~~~warm absorber fit} & \multicolumn{2}{l}{references} & comments \\
        &  & $\Gamma_{\rm x}^{\rm intr}$ & log $U$~ & log $N_{\rm w}$ & sug. & mod. & \\
      \noalign{\smallskip}
  \hline
  \hline
  \noalign{\smallskip}
    {3C\,212}  &  `red' quasar &      &       &   & [1] & & dusty WA not yet fit \\
     ~~~df     &   &  --2.4  & --0.5  & 21.9  & [1] & & best-fit parameters of dust-free WA \\
    {IRAS\,13349+2438} & NL quasar & --2.9 & --0.4 & 21.2$^{1}$ & [2], [3] & [10], [11] & \\
     ~~~df  & & --2.2 &  ~0.7 & 22.7 & & & dust-free WA for comparison\\
    {NGC\,3227}  & Sy 1.5 &--1.9 & --0.3 & 21.8 & [4], [5] & [4] & \\
    {NGC\,3786}  & Sy 1.8 &--1.9 & --0.8 & 21.7 & [6] & [6] & \\
    {MCG\,6-30-15} & Sy 1 &--2.2&       & 21.7 & [7] & [7] & ($>$)2-comp. WA  \\
    {IRAS\,17020+4544} & NLSy\,1 & --2.8 & ~0.7 & 21.6$^{1}$ & [8] & [12] \\
     ~~~sil & & --2.4 & ~1.0 & 21.6$^{1}$ &  & [12] & silicate species only\\
     {4C\,+74.26} & radio quasar & --2.2 & --0.1 & 21.6 & [9] & [9] & excess cold absorption possible \\
  \noalign{\smallskip}
  \hline
     \end{tabular}
  \label{tab1}

  \noindent{ \scriptsize $^{(1)}$ fixed to
     the value $N_{\rm opt}$ determined from optical reddening.
   References: [1] Mathur 1994, [2,3] Brandt et al. 1996, 1997,
      [4] Komossa \& Fink 1997b (see also Komossa \& Fink 1996),
      [5] George et al. 1998b, [6] Komossa \& Fink 1997c, [7] Reynolds et al. 1997
          (the possibility of a dusty WA in this galaxy was 
           earlier briefly mentioned in Reynolds 1997),
      [8] Leighly et al. 1997, [9] Komossa \& Fink 1998,
      [10] Komossa \& Greiner 1999, Komossa et al. 1999b, [11] Siebert et al. 1999,
      [12] Komossa \& Bade 1998.
     }
\end{table*}

\section{Dusty warm absorbers}

Below, we describe in more detail the properties of dusty warm material,
and provide a summary of the suggested candidates for dusty WAs
including explicit spectral model fits. 

\subsection{Modifications of the X-ray absorption structure}

Several changes occur in the presence
of {\em dust} mixed with the warm absorber:
(i) Gas-dust interactions influence the thermal conditions in the gas and change
its ionization state (e.g., Draine \& Salpeter 1979). In particular,
for high ionization parameters dust effectively competes with the gas in absorbing
photons (Laor \& Draine 1993).
(ii) Dust scatters and absorbs the incident radiation  
and X-ray absorption edges are created by
inner-shell photoionization of metals bound in the 
dust (see Figs 4,5 of Martin \& Rouleau 1991).

Under the following assumptions we (Komossa \& Fink 1997b,
Komossa \& Bade 1998) have calculated
a sequence of dusty warm absorber models using
Ferland's (1993) code {\em Cloudy}: 
(i) The dust composition (graphite and astronomical silicate)
and grain size distribution were chosen like
in the Galactic diffuse interstellar medium (Mathis et al. 1977)
if not mentioned otherwise, as incorporated in
{\em Cloudy}, 
(ii) the metal abundances
were depleted correspondingly (a mean of Cowie \& Songaila 1986;
see Ferland 1993, Baldwin et al. 1991 for details), 
(iii) the warm material was assumed to be of constant
density and ionized by a `mean Seyfert' IR to gamma-ray continuum
(with $\alpha_{\rm uv-x}$=--1.4 in the EUV). 

The main consequences of the presence of dust internal
to the warm absorber can be summarized as follows:
\begin{itemize}
\item Absorption edges in the X-ray spectrum
from neutral metals bound in the dust, like a carbon CI
edge (from the graphite species) and an oxygen OI edge
(from silicate) are predicted. [The shift in edge energy due to
solid state effects (e.g., Martin 1970, Adamietz et al. 1993) is only of the order
of a few eV (Greaves et al. 1984).] 

\item There is a stronger temperature gradient across the absorber, 
with more gas in a colder state. This and the previous effects lead
to an {\em effective flattening} of the X-ray spectrum
(Fig. 2) 

\item 
The increased sensitivity of dust to radiation pressure 
(e.g., Chang et al. 1987, Binette et al. 1997) 
likely causes the gas to outflow (if not otherwise confined) 
leading to temporal
changes of the absorber properties. 
[Indeed, strong variability in X-ray count rate of factors 10-15
has been detected from the dusty-warm-absorber candidates
NGC 3227 and NGC 3786 (Komossa \& Fink 1997b,c.)] 
 
\item The warm absorber emission in certain optical/UV lines, 
in particular the iron coronal lines, is much reduced
because of the depletion of iron into dust grains.  
[This likely also alleviates the problem of the 
recently reported potential over-prediction
of these lines from some warm absorbers; see Sect. 4 below].  
\end{itemize} 

\subsection{Results from X-ray spectral fits: List of suggested/modeled candidates} 

We have investigated most suggested dusty-WA candidates 
with our models. We get successful X-ray spectral
fits. In some cases the best-fit parameters can deviate
substantially from those obtained by fitting a 
dust-{\em free} warm absorber, though.
Table 1 gives a summary of the suggested dusty warm absorber
candidates and the model results.    
The ionization parameters are found to be somewhat lower 
than those of dust-free warm absorbers, with the exception of 
IRAS\,17020.
It is interesting to note that the dusty warm absorbers 
show a rather narrow range in column densities $N_{\rm w}$
which differ by less than a factor $\sim$2. This is likely
partly caused by selection effects: 
Firstly, X-ray warm absorbers of {\sl lower column density} are difficult
to detect presently, and small deviations of the Balmer decrement 
from the recombination value are less straight forwardly
interpreted in terms of reddening. Secondly, in case of 
{\sl higher} column density the dusty gas becomes optically thick,
and the concept of the standard extinction curve can no longer be applied.

\subsection{Origin of the dusty material}

In case dust is mixed with the warm absorber, important
constraints on its density and locations can be
obtained from the requirement of dust survival:
Firstly, dust is heated and destroyed by the 
radiation field of the central source if
at distances less than 
 $r_{\rm ev} \approx \sqrt{L_{46}}$ 
(e.g., Netzer 1990; see Laor \& Draine 1993 for a dependence
of this relation on grain size).  
Secondly, dust is heated by collisions with
gas particles (e.g., Baldwin et al. 1991)
to a temperature that exceeds the evaporation temperature
if the gas density is high enough.  
Both approaches locate the dusty warm absorbers
outside the BLR.

One component that suggests itself to be identified
with the dusty warm absorber is an atmosphere of the 
molecular torus 
which plays an important role in unification schemes. 
The torus has the advantage that it is `known' to be present
(no need to invent a new component), and would provide
a natural reservoir for dust close to the central power source 
(but far enough away to ensure dust survival). 
One is then lead to a viewing geometry such that
our line of sight grazes the torus atmosphere in those
Seyferts that show dusty warm absorbers (see also the discussion
in Brandt et al. 1996, Reynolds 1997, and Komossa \& Fink 1997b,c). 

Looking at Tab. 1, one question arises immediately: 
the dusty warm absorbers seem to come in {\em all types} of
Seyfert galaxies, which is not what would be expected
in a simple viewing-geometry-scenario. 
The sample is still small, though, and it might be
interesting to re-investigate this possibility once
the dusty warm absorbers are confirmed, and further ones
are known.

The other already earlier suggested dusty component of the 
active nucleus, and thus
potential dusty warm absorber counterpart, is the 
dusty transition region between BLR and NLR `invented' and studied
by Netzer \& Laor (1993).  
The ionization parameter of this region should show
a smooth transition to that of the BLR and NLR, though;
whereas the ionization parameters of warm absorbers are
typically much higher.

\subsection{Model uncertainties}

The line of arguments in favor of dusty warm absorbers 
(reviewed in Sect. 2.2.2; see Brandt et al. 1996 and Komossa \& Fink 1997b
for many further details) relies on several assumption. 
Firstly, optical and X-ray observations were not performed
simultaneously. Variability in the optical reddening
on the timescale of years is possible in Seyferts (e.g., Goodrich 1989, 1995), 
but we note that for both, IRAS\,17020 and IRAS\,13349
optical spectra taken at many different epochs are
available and they do not show evidence for 
changes in reddening (e.g., Siebert et al. 1999). 
Secondly, optical-UV and X-ray photons may not travel
along the same l.o.s, and they may therefore `see'
different amounts of dust and gas. This would be the case if the X-rays
were seen mainly in scattered light, or if there 
was a strong starburst contribution. Since most
of the Seyferts with dusty warm absorber candidates are rapidly
variable in X-rays (e.g., a factor of $\sim$2 within 800 s
and a factor $\sim$10 within a few years in case of NGC\,3227;
Komossa \& Fink 1997b) these two possibilities can
be excluded.

As far as the {\sl modeling} of dusty warm absorbers is concerned,
one uncertainty is the applicability of the code {\em Cloudy}. 
More precisely, radiation pressure
may drive strong outflows of the dusty gas.
In case of large velocity gradients, the treatment of
overlapping resonance lines becomes inaccurate (Ferland 1993),
and the heating-cooling balance might be disturbed by
expansional cooling of the gas.

Another uncertainty are the dust properties: Is the
dust in Seyfert galaxies sufficiently similar to
Galactic ISM-like dust (and is the latter really `MRN'-like
in nature{\footnote{see, e.g., Laor \& Draine 1993, and the 
recent review by Witt 1999}}) ? 
Firstly, galactic ISM-like dust was also assumed  
to estimate the amount of dust from the
optical spectra. Therefore, making the same assumption in
modeling the X-ray data is a good starting point. 
Secondly, observations by ISO suggest that the 
dust/gas ratio in other galaxies does not deviate 
much from the Galactic value (e.g., Bianchi et al. 1999). 
Thirdly, even for other dust compositions or structures,
the X-ray absorption structure would not change strongly,
because the X-ray spectrum is not dominated by complex
molecular transitions and solid state and chemical effects
(these may have some influence on the heating-cooling balance,
though), but by K-shell 
photoionization of metals bound in the dust. 

\subsection{Future perspectives} 

Given the presence of X-ray absorption edges of neutral 
dust-bound metals like the carbon edge,
dusty warm absorbers, if  
confirmed{\footnote{Even if the dusty X-ray warm absorbers are not confirmed,
this likely tells us a lot about the dust in these galaxies,
which then would have rather different properties from Galactic
ISM-like dust, namely a different size distribution with
a dominance of small grains (which more effectively scatter, thus more effectively
redden the optical/UV spectrum), or a very different gas/dust ratio !}}
with {\sl XMM}, {\sl AXAF} and {\sl ASTRO-E},
will play an important role
not only in probing components of the active nucleus, like the dusty torus,
but also are they a very useful diagnostic 
of the (otherwise difficult to determine)
dust properties
and dust composition in other galaxies.

%
\begin{table*}[t]
\caption {Predicted iron lines (Komossa \& Fink 1997a) compared
to the observed ones of NGC\,4051 assuming a covering factor
of the warm absorber of 50$\%$; all lines are
given relative to the {\em observed} H$\beta$ luminosity of
NGC\,4051.
Some notes: (i) The strongest predicted
among the Fe lines in most models is Fe\,19\,$\lambda$1118; (ii) model results 
are robust
against the small uncertainties in the reddening correction (e.g., we assumed
A$_{\rm v}$ = 0.7mag, Nagao et al. (2000) used A$_{\rm v}$ = 1mag); (iii) all non-`standard'
model sequences were first re-fit to the data, to derive the {\em new} best-fit
WA parameters (which sometimes slightly differed from the old ones) before the
lines were predicted [if this is not done, 
the predicted Fe lines change more strongly!], 
(iv) recent NLR(and torus-)related multi-component photoionization models of Sy\,1s can
successfully account for [FeVII] (e.g., Komossa \& Schulz 1997, Murayama \& Taniguchi 1998a,b).
 }
 \begin{tabular}{lccl}
\hline
\noalign{\smallskip}
model & line predictions & & comments \\
      & [FeVII],[FeX],Fe[XI]/H$\beta_{\rm obs}$ & [Fe XIV]/H$\beta_{\rm obs}$ & \\
\noalign{\smallskip}
\hline
\hline
\noalign{\smallskip}
`standard' & $<$5\,10$^{-5}$, $\ll$\,obs. & 0.004 & {\small observed: $Fe14/H\beta<0.06$} (Nagao et al. 2000) \\
\hline
\noalign{\smallskip}
var. abundances & $''$ &  {\small change $\le$ fact.\,2-4} & $0.3 \le Z \le 1$\\
 & & & HeII$\lambda$4686 becomes strongest opt. line\\
\hline
\noalign{\smallskip}
var. density & $''$ & $''$ & 4 $\le \log n_{\rm H} \le 9.5$  \\
 & & &  \\
\hline
\noalign{\smallskip}
var. IR conti. & $''$ & $''$ & up to the observed IR data points (from host \\
 & & & galaxy) $\rightarrow$ increased f-f heating  \\
\hline
\noalign{\smallskip}
var. EUV conti. &  $''$ &  $''$ & additional soft excess to account for $L_{\rm H\beta}$ \\
 & & & $\log T_{\rm bb}=5, Q_{\rm bb}=Q_{\rm pl}$ \\
\hline
\noalign{\smallskip}
`91 McHardy data & $''$ & 0.02 & $\log U = 0.2$, $\log N_{\rm w} = 22.47$ (our re-fit of the data \\
 & & & originally presented by McHardy et al. 1995) \\
\noalign{\smallskip}
\hline
\end{tabular}
\end{table*}

\section{The warm absorber in NGC\,4051, and the (missing) connection between WA and CLR}

\subsection{Introduction} 

Given the several good candidates for dusty warm absorbers
the question emerges whether {\em all} warm absorbers contain 
dust. 
We think this is unlikely because (i) warm absorbers
have also been observed in several Seyfert\,1 galaxies that do not
show evidence for optical reddening, and (ii) in some cases
the modifications in the X-ray absorption spectrum due to  
the presence of dust are so strong
that no successful X-ray spectral fit is possible.   

Dust-free warm absorbers might contribute significantly 
to the emission in optical/UV iron coronal lines.
In the following we shall examine in detail the warm absorber
in the NLSy1 galaxy NGC\,4051 with respect to the question 
whether this ionized material can account for the 
optical coronal lines observed in this galaxy.
This is an update of our earlier work on this subject
(see Komossa \& Fink 1997a; KF97a hereafter). 

\subsection{Observed properties of NGC\,4051} 

We first give a brief summary of previous observational
properties of NGC\,4051 relevant for the present study (taken from KF97):
(i) To model the warm absorber and to predict its line emission,
we compiled the observed SED of the galaxy from
the literature (see Fig. 1 of KF97a); based on
photon counting arguments, we found evidence for an
EUV bump in the spectrum of NGC\,4051.
(ii) Intensity ratios of coronal lines were taken from
the literature (e.g., de Robertis \& Osterbrock 1984, 
Penston et al. 1984, Peterson et al. 1985,
Wilson \& Nath 1990, Osterbrock et al. 1990, Erkens et al. 1997, 
Nagao et al. 2000).   
(iii) A warm absorber fit to our {\sl ROSAT} PSPC spectrum
of NGC\,4051 yields $\log U$ = 0.4 and 
$\log N_{\rm w}$ = 22.7.{\footnote{We have explored the possibility
that the warm absorber in NGC\,4051 contains dust. 
The optical spectrum of NGC\,4051 is not heavily reddened,
but the Balmer decrement of broad H$\alpha$/H$\beta$ deviates
from the recombination value. If we try to model
the X-ray spectrum of NGC\,4051 with a one-component
dusty warm absorber we do not get a successful spectral
fit. The possibility remains, though, that there is
a second, dusty warm absorber of low column density
present. In fact, there are some residuals around the Carbon 
edge after fitting a dust-free warm absorber. Since these
could also be of instrumental origin, higher spectral
resolution data are required to search for a second, 
dusty, warm absorber in NGC\,4051.}}
Further, from variability arguments we derive a limit on the 
density of the warm absorber, $n_{\rm{H}} \approxlt$ $3 \times 10^{7}$cm$^{-3}$,
which translates into a distance from the 
nucleus of $r \approxgt 3 \times 10^{16}$ cm.    

\subsection{Prediction of coronal lines} 

We then checked, for this best-fit warm absorber model, 
the strengths of the optical/UV lines originating from
the ionized material; in particular, the Fe lines
[FeVII]$\lambda$6087, [FeX]$\lambda$6374, [FeXI]$\lambda$7892,
and [FeXIV]$\lambda$5303 (see Tab. 2).
We find that all of them are {\em much weaker} than observed!

We then varied those parameters that not strongly
influence the X-ray absorption structure, but
could have an effect on the predicted Fe line strengths;
namely: the metal abundances, the gas density, the EUV continuum
shape, the IR continuum strength. 
We find that in all cases, the lines [FeVII]--[FeXI] remain 
much weaker than observed by several orders of
magnitude.
The reason for this is that the warm absorber is 
always {\em too highly ionized}, with a totally
negligible amount of Fe$^{9+}$ and Fe$^{10+}$ ions in the gas.
Therefore, changes in collisional strengths for 
the relevant Fe transitions, which are still poorly known,
are not expected to alter this result. 
We conclude that, for the case of NGC\,4051,
{\em warm absorber and coronal line region are {\sl not}
one and the same component, but of different origin}.  
This is consistent with the recent findings of Nagao
et al. (2000) that the [FeXI] emission of NGC\,4051 is not
confined to the nucleus, but widely
extended (out to at least $\sim$150 pc). 

\subsection{Comparison with subsequent studies} 

Recently Porquet et al. (1999; P99 hereafter) presented a parameter space study
of the strengths of Fe coronal lines that originate
from warm absorbers. They conclude that Fe lines
in low-density absorbers (they studied the density range
$\log n_{\rm H}$ = 8--12) are over-predicted for part of the parameter space.
Comparing our earlier results on NGC\,4051 (Komossa \& Fink 1997a) 
with their recent results, we find both to be consistent:  
For warm absorbers dominated by OVIII absorption and high
ionization parameters, no overprediction in line emission
occurs (their Tab. 1).
 
The question remains whether the `OVII absorbers' of P99 do
indeed overpredict the Fe lines and thus are in conflict  
with observations. 
We suggest that most of the strong OVII absorbers likely contain dust
(which was not included in the models of P99), as 
proposed by Reynolds (1997). 
The strong depletion of Fe into dust grains then
results in weaker gas phase emission in the Fe coronal
lines.   

\subsection{A warm absorber relation to any optical/UV emission line component ?}

Finally, we note that our more general search for a warm-absorber origin of
of one of the emission-line regions observed
in the UV/optical spectra of Seyfert galaxies did
not give a single example with a one-to-one match
(see Komossa \& Fink 1997e for a brief overview);
emission lines from warm absorbers are always {\em weaker}
than observed lines.
However, the possibility of a (weak) warm-absorber
contribution to individual emission lines (e.g., CIV
in the case of NGC\,3227; Komossa \& Fink 1997b)
remains.

\section{The warm absorber of MR\,2251--178} 

\subsection{Introduction}

MR\,2251-178 at redshift $z$=0.068 
was the first quasar to be discovered in X-rays, in the course
of the {\sl Ariel V} all-sky survey
(Cooke et al. 1978, Ricker et al. 1978).   
It turned out to be an outstanding
object in many respects. It has a high ratio of $L_{\rm x}/L_{\rm opt}$,
is surrounded by a giant
HII envelope observed in [OIII] emission, 
and was the first object in which a warm absorber was 
discovered (Halpern 1984).  
Based on {\sl EXOSAT} and {\sl Ginga} observations 
Pan et al. (1990) and Mineo \& Stewart (1993), again, 
interpreted and modeled the data in terms of the presence of a warm absorber.
In contrast, Walter \& Courvoisier (1992), in an independent
analysis explained the observations in terms of a
variable powerlaw spectral shape, and concluded the
presence of a warm absorber was unnecessary to explain
the X-ray data. Reynolds (1997), in the course of a large sample
study of {\sl ASCA} observations of AGN reported the detection
of OVII and OVIII absorption edges in the X-ray spectrum of MR\,2251.  

Here, we present first results from a detailed
analysis of a deep archival {\sl ROSAT} PSPC 
(Tr\"umper 1983, Briel et al. 1994) observation of 
this source performed in Nov. 1993 with a duration
of 18 ksec. 

\subsection{X-ray analysis: results}

\begin{figure}[t]
\hspace*{1.0cm}
      \vbox{\psfig{figure=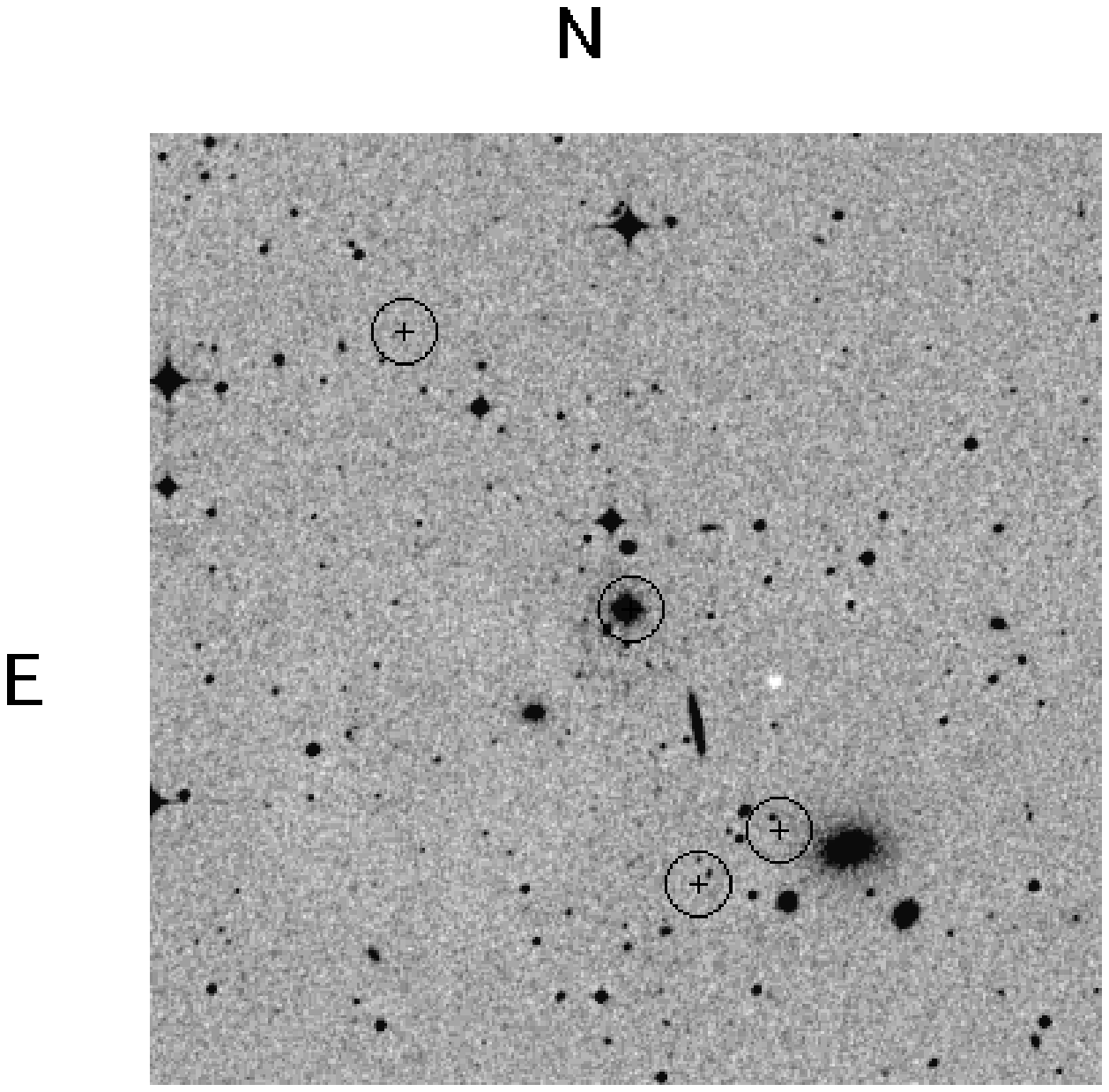,width=6.7cm,%
          bbllx=5.7cm,bblly=13.2cm,bburx=16.1cm,bbury=23.5cm,clip=}}\par
 \caption[mr_ima]{X-ray sources detected
in a 10\arcmin~$\times$ 10\arcmin~field
around MR\,2251-178 superimposed on an optical image. 
The circles drawn around
the X-ray source positions are of 20\arcsec~ radius.
   }
 \label{mr_ima}
\end{figure}

X-ray data reduction was carried out
in a standard manner (see Komossa 2000, in prep.).
In total, 36~X-ray sources were detected with 
a likelihood $\ge$ 10 within the field of view.
The positions of those in the vicinity of MR\,2251 
are shown in Fig. \ref{mr_ima}, overlaid
on an optical image from the digitized Palomar sky survey.

\begin{figure}[ht]
      \vbox{\psfig{figure=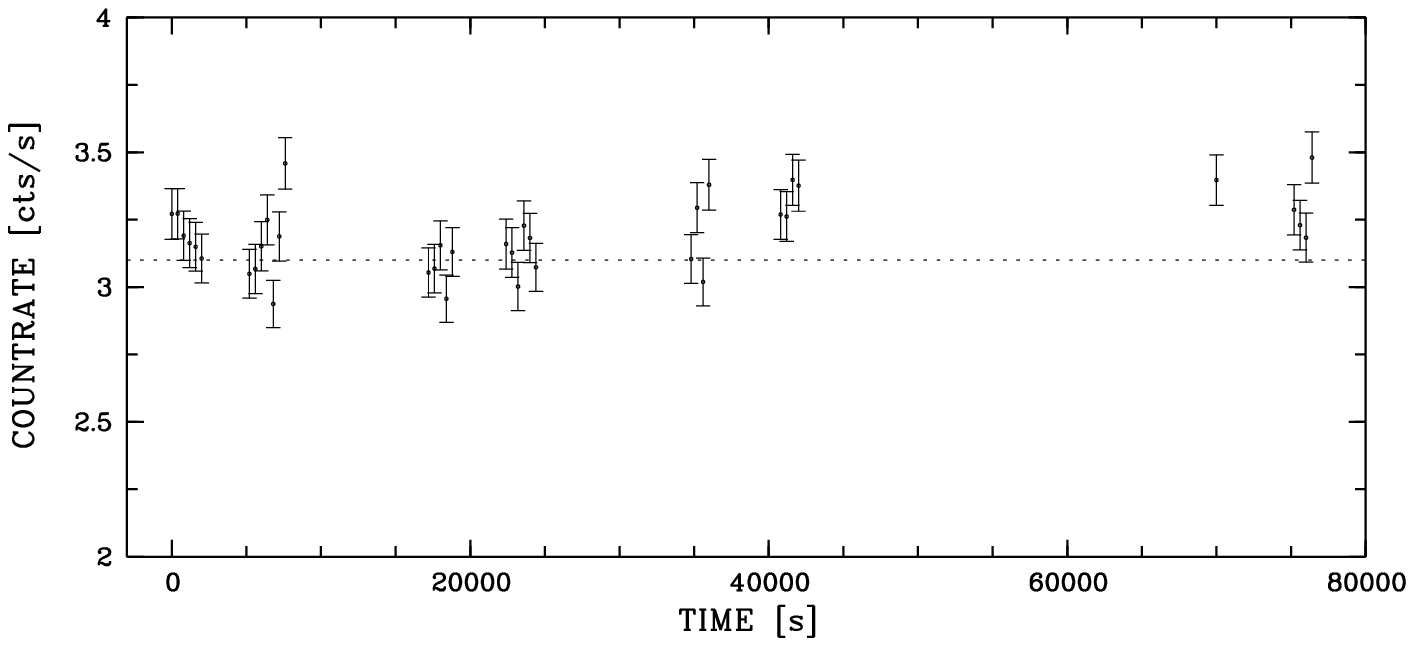,width=8.8cm,%
         bbllx=2.85cm,bblly=3cm,bburx=18.1cm,bbury=11.3cm,clip=}}\par
  \vspace{-0.3cm}
\caption[mr2251_light]{X-ray lightcurve of MR\,2251-178, binned to time intervals of 400 s. The time
is measured in seconds from the start of the pointed observation. The mean countrate
during the RASS, not shown, was 1 cts/s. 
 }
\label{mr2251_light}
\hspace*{0.7cm}
      \vbox{\psfig{figure=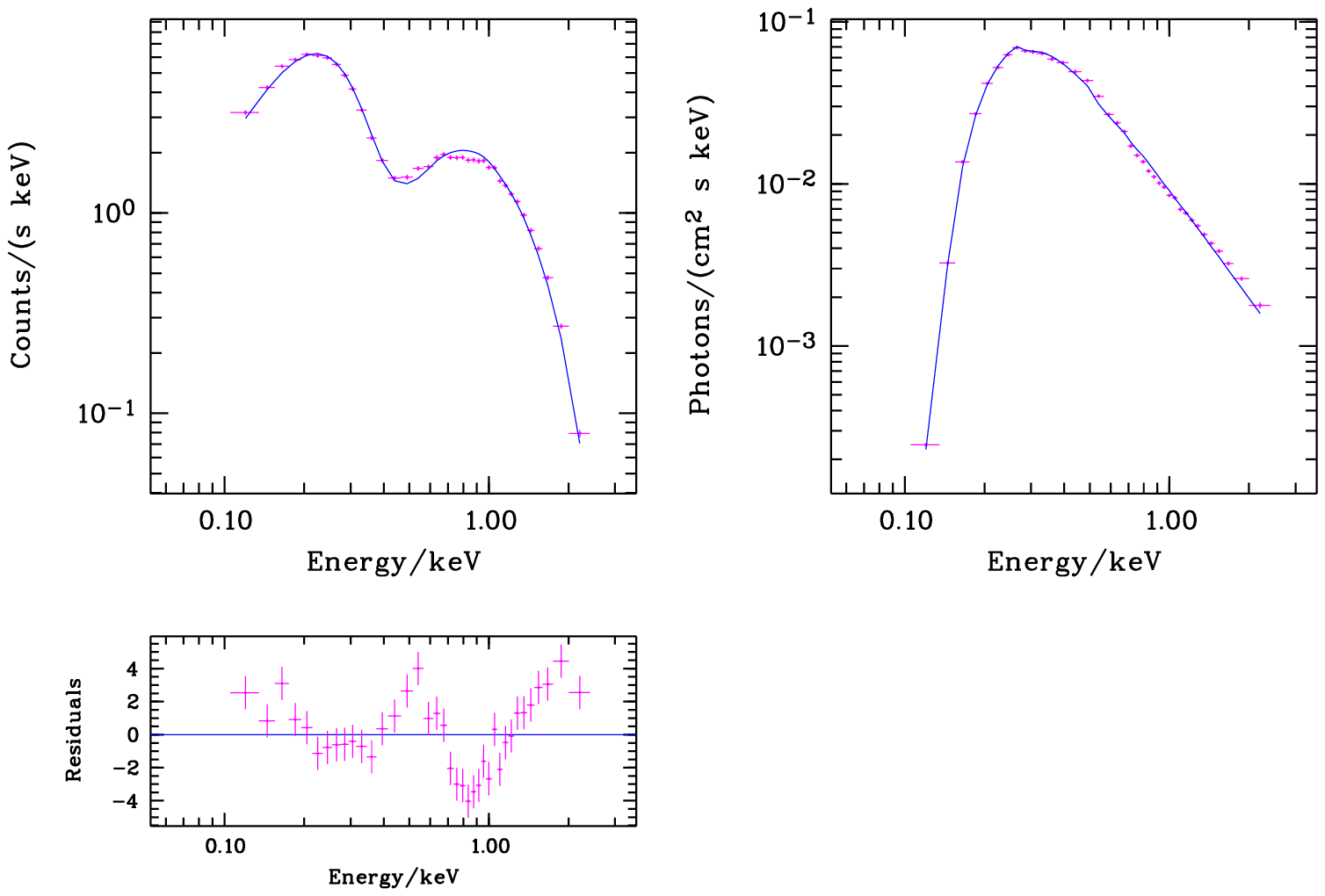,width=6.4cm,%
          bbllx=2.5cm,bblly=5.8cm,bburx=10.1cm,bbury=11.9cm,clip=}}\par
\hspace*{0.7cm}
      \vbox{\psfig{figure=komossa2_fig7a.ps,width=6.4cm,%
          bbllx=2.5cm,bblly=1.25cm,bburx=10.1cm,bbury=4.4cm,clip=}}\par
\hspace*{0.7cm}
      \vbox{\psfig{figure=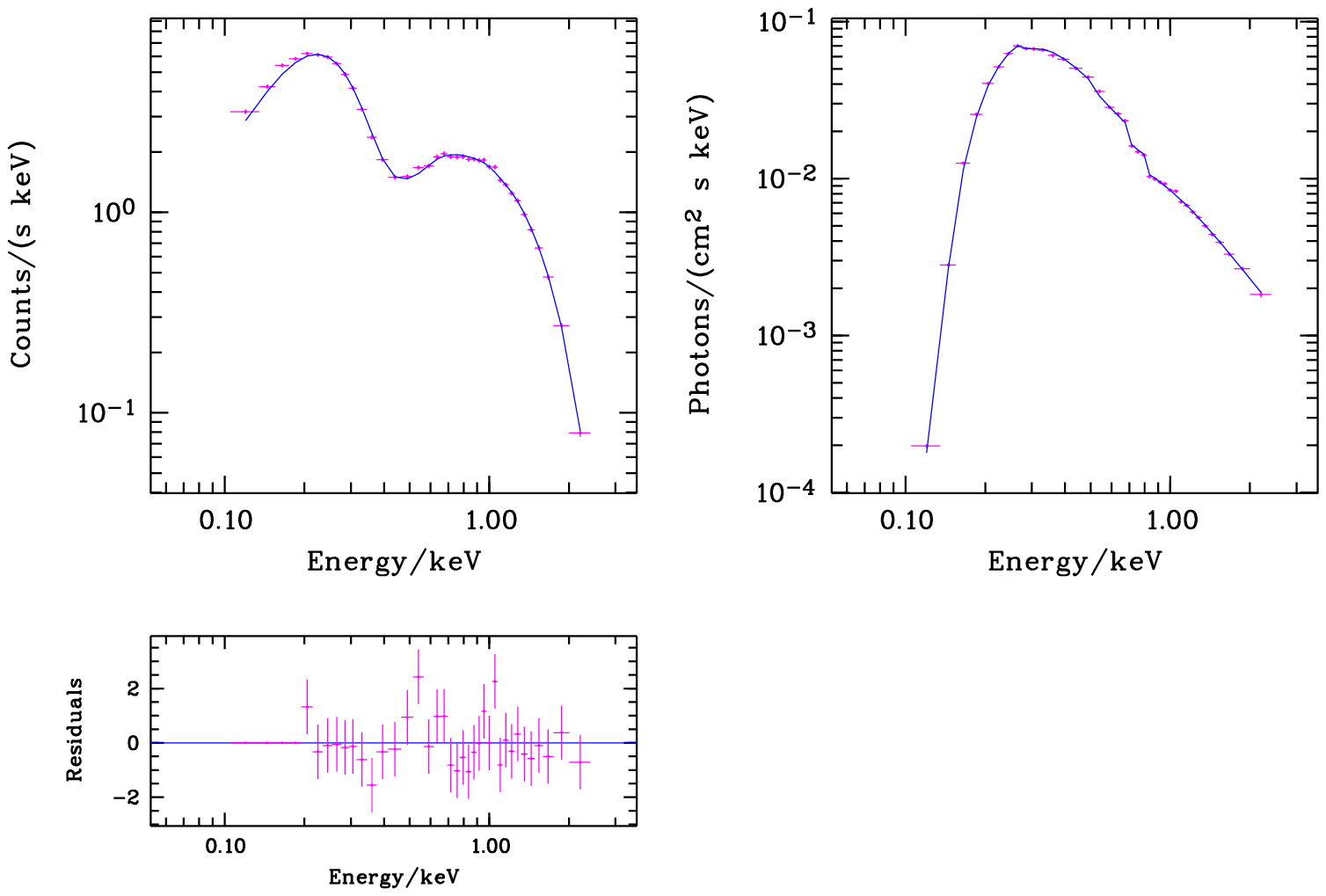,width=6.4cm,%
          bbllx=2.5cm,bblly=1.25cm,bburx=10.1cm,bbury=4.4cm,clip=}}\par
 \vspace{-0.2cm}
\caption[mr_x]{The upper panel shows the observed X-ray spectrum of MR\,2251 (binned
to S/N=35; crosses)
and the best-fit powerlaw model (solid line).    
The second panel displays
the fit residuals for this model, whereas the lowest panel gives
the residuals from a warm absorber fit parameterized by two absorption edges
(note the different scale of the ordinate).
The lowest energy bins were ignored in this fit (see text for details).  
  }
\label{mr_x}
\end{figure}

The mean source countrate was 3.1 cts/s, a factor
of 3 stronger than during the {\sl ROSAT} all-sky
survey observation performed in 1990.
The X-ray lightcurve is displayed in Fig. \ref{mr2251_light}.  

First, we fit a single powerlaw to the total X-ray spectrum.
Cold absorption was fixed to the Galactic value towards MR\,2251,
$N_{\rm Gal} = 2.77 \times 10^{20}$ cm$^{-2}$ (Lockman \& Savage 1995),
since it is underpredicted otherwise.  
This gives $\Gamma_{\rm x}=-2.3$ and clearly is a very poor description of the
data ($\chi^2_{\rm red}=5.4$; Fig. \ref{mr_x}).
A Raymond-Smith emission model does not fit the spectrum
at all, even if temperature, normalization, metal abundances
and the amount of cold absorption are all treated as free 
parameters.   
We then tried two-component spectral fits involving a powerlaw
plus soft excess parametrized by different models. The 
quality of the fit remains unacceptable, though.    

The presence of a warm absorber markedly improves
the fit. Performing a two-edges fit with edge energies fixed
at the theoretical values of OVII and OVIII we obtain
$\tau_{\rm OVII}=0.26\pm{0.12}$, $\tau_{\rm OVIII}=0.20\pm{0.12}$,
and $\Gamma_{\rm x} = -2.20\pm{0.02}$. 
The not yet totally satisfactory quality of the fit,
$\chi^2_{\rm red}=1.6$, can be traced back to a deviation of
the low energy part of the spectrum (between 0.1--0.18 keV)
from the model prediction which already caused an underprediction
of the galactic $N_{\rm H}$ in the pure powerlaw fit.
If this part of the spectrum is excluded from the spectral
fitting, we obtain $\chi^2_{\rm red}=0.9$, $\tau_{\rm OVII}=0.22\pm{0.11}$,
$\tau_{\rm OVIII}=0.24\pm{0.12}$, and $\Gamma_{\rm x} = -2.21\pm{0.02}$.
We conclude that a warm absorber is clearly present in this
source.

\section{The X-ray transient AGN RXJ0134-4258} 

\subsection{Summary of the observations} 

The Narrow-line Seyfert\,1 galaxy 
RXJ0134-4258 is one of the rare sources which showed 
dramatic spectral variability. Its spectrum
changed from ultrasoft ($\Gamma_{\rm x}=-4.3$) in the {\sl ROSAT}
all-sky survey (RASS) to flat ($\Gamma_{\rm x}=-2.2$)
in our pointed
PSPC observation taken two years later while
its count\-rate remained nearly constant (Greiner 1996,
Komossa \& Fink 1997d, Komossa \& Greiner 1999,
Komossa et al. 1999b, Grupe et al. 2000, Komossa \& Meerschweinchen 2000).  

One possible explanation for this kind of spectral
variability is the presence of a warm absorber. 
In particular, we have studied the following two scenarios
(details are given in Komossa \& Meerschweinchen 2000; KM2000 hereafter):
\subsection{Explanations invoking a warm absorber} 

\noindent{\em {\underline{Changes in ionization state of a warm absorber:}}}
We find that a warm absorber of ionization parameter
$\log U \simeq 0.5$ and column density $\log N_{\rm w} \simeq 23$ fits
well the RASS observation of this source ($\chi^2_{\rm red}=0.6$; Fig. 8).
To account then for the much flatter spectrum during the later
PSPC observation requires a change in ionization state of the warm
absorber. Since the intrinsic luminosity of the source did not
vary strongly from one to the other observation, it is then
required that the ionization state of the warm absorber reflects
the (unobserved) history of the variability in the intrinsic luminosity
(see KM2000 for details).
After allowing for non-equilibrium effects in the absorber and/or a range
in densities, such a warm absorber is consistent with the long-
and short-timescale variability behavior of RXJ0134-4258.
We did not favor this possibility in our earlier work, because
it introduces a new level of complexity (more free parameters) as
compared to the simpler case of an absorber in equilibrium.

\begin{figure}[th]
\hspace*{0.1cm}
\psfig{file=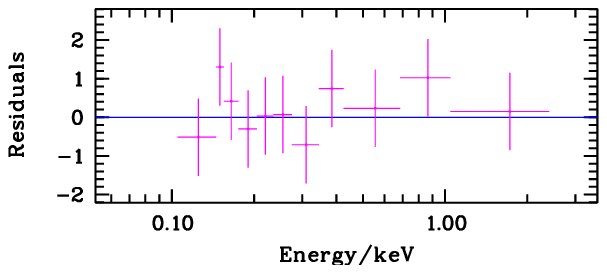,width=6.7cm,clip=} 
\caption[surv_0134]{Residuals after fitting a {\em warm absorber} to the
{\em RASS} spectrum (= `steep-state' spectrum) of RXJ0134-4258. 
}
\vspace*{0.3cm} 
      \vbox{\psfig{figure=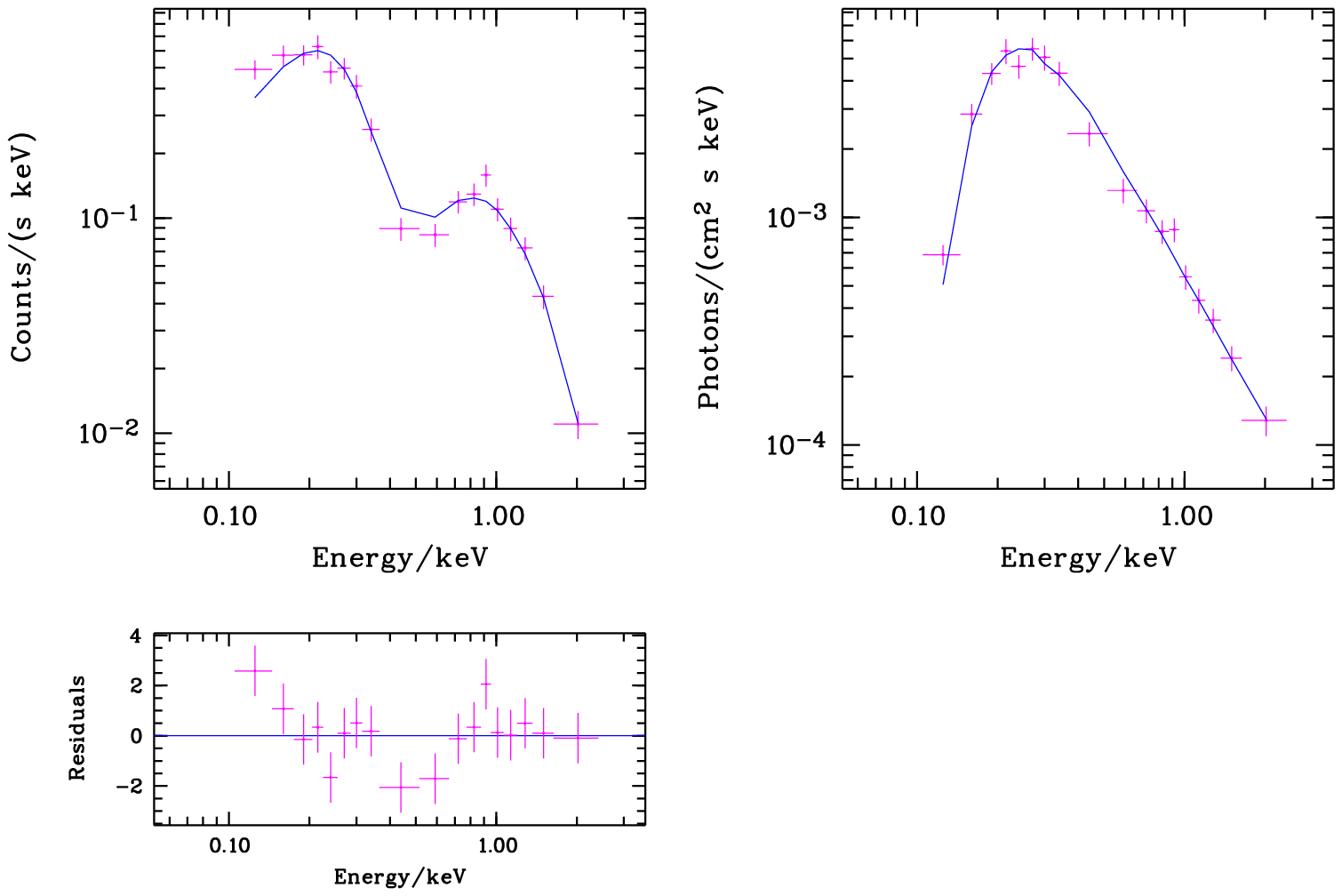,width=6.9cm,%
          bbllx=2.5cm,bblly=1.4cm,bburx=10.1cm,bbury=4.4cm,clip=}}\par
\caption[poi_0134]{Residuals after fitting a {\em powerlaw} to the
{\em pointed} PSPC data (= `flat-state' spectrum) of RXJ0134-4258
(see also footnote 7). 
}
\end{figure}
 
\noindent{\em{\underline{Cloud passing our line-of-sight:}}} 
Alternatively, a cloud of ionized material may have passed
our line of sight during the RASS observation, 
and has (nearly) disappeared during the later PSPC observation 
(see KM2000 for details). 

In summary, we find that a warm absorber fits well the {\sl ROSAT} survey
observation, and residuals around the expected location
of absorption edges are still present during our later
pointed PSPC observation.{\footnote{The flat-state spectrum 
shows two kinds of residuals when fit by a single powerlaw
(cf Fig. 9):
Deviations of the lowest energy bins near 0.1 keV suggesting
the presence of a very soft excess (or instrumental features; 
they are similar to the ones seen in MR\,2251, Sect. 5.2), 
and deviations near the
expected locations of the WA absorption edges. No single-component
fit can remove all residuals.  The simultaneous presence
of both, a soft excess and a warm absorber is well possible, 
but such fits would be an over-representation of the PSPC
spectral resolution. Since both features, very soft excess and
warm absorber (note the redshift of RXJ1242) are outside or
very close to the border of the {\sl ASCA} sensitivity range,
further observations with the new generation X-ray
telescopes are required to search for these features.}}   
In addition, the presence of high-ionization UV absorption 
lines in this object was recently reported by Goodrich
(talk at the Narrow-line Seyfert\,1 workshop in Bad Honnef, Dec. 1999),
and the recent study by Crenshaw et al. (1999) suggests a near
one-to-one match of the presence of UV and X-ray warm absorbers.  

We conclude, that the presence of a time-variable warm absorber is
a viable explanation for all available X-ray observations
of RXJ0134-4258. 
High-spectral resolution data taken
during episodes of repeated spectral variability of this source
are required to distinguish this from other possible scenarios
like real changes in the steepness of the powerlaw or a 
powerlaw plus soft excess description (see Komossa \& Meerschweinchen 2000).   

\begin{acknowledgements}
It is a pleasure to thank Gary Ferland for providing {\em Cloudy},
Smita Mathur for very useful and pleasant discussions, and the organizers
for carrying out this exciting workshop. 
This workshop was funded by the Inter-Research Centers Cooperative Program
of the JSPS and the Deutsche Forschungsgemeinschaft DFG.   
The \ros project has been supported by the German Bundes\-mini\-ste\-rium
f\"ur Bildung, Forschung und Technologie
(BMBF/DLR) and the Max-Planck-Society. \\ 
Preprints of this and related papers can be retrieved from our webpage
at http://www.xray.mpe.mpg.de/$\sim$skomossa/
\end{acknowledgements}

\end{document}